# Experimental evidence for a second hydrogen ordered phase of ice VI

Tobias M. Gasser[a], Alexander Thoeny[a], Lucie Plaga[b], Karsten W. Köster[b], Martin Etter[c], Roland Böhmer[b], Thomas Loerting[a],*

[a]Institute of Physical Chemistry, University of Innsbruck, 6020 Innsbruck, Austria

[b] Fakultät Physik, Technische Universität Dortmund, D-44221 Dortmund, Germany

[c] Deutsches Elektronen-Synchrotron (DESY), 22607 Hamburg, Germany

e-mail: thomas.loerting@uibk.ac.at

## Abstract

In the last decade five new ice phases were experimentally prepared. Two of them are empty clathrate hydrates and three of them represent hydrogen ordered counterparts of previously known disordered ice phases. Here, we report on hydrogen ordering in ice VI samples upon cooling at pressures up to 1.8 GPa. Using calorimetry, dielectric relaxation spectroscopy, Raman spectroscopy, and powder X-ray diffraction we provide evidence for the existence of a second hydrogen ordered phase related to ice VI, that we call ice β-XV. This phase is more ordered than ice XV by 14% and directly transforms to ice XV above 103K and to ice VI above 129K. That is, upon heating an order➜order➜disorder transition is experienced. The new phase is thus thermodynamically more stable than ice XV requiring a new stability region in the phase diagram of water. Raman spectroscopy indicates ice XV and ice β-XV to be different in terms of symmetry and space group. The activation energies for dielectric relaxation are 45 kJ mol$^{-1}$ in ice β-XV compared to 18 kJ mol$^{-1}$ in ice XV. Powder X-ray data show the oxygen network to be the one of ice VI. The ordering of hydrogen atoms induces a significant peak shift to lower d-spacings at d=0.265 nm in ice β-XV, whereas for ice XV shifts to higher d-spacings are found. This present work represents a unique experimental realization of a second electric ordering in an ice phase.

## Introduction

One of water´s anomalies is its rich polymorphism. Currently 17 crystalline phases are known.(1) These polymorphs can be categorized as hydrogen ordered and hydrogen disordered, where typically the disordered form has exactly one ordered pendant. Some ordered variants, such as for ice IV, are awaiting their discovery(2), and also no experimentally confirmed examples are known for hydrogen disordered phases with more than one ordered counterpart. This is surprising since many different ways of ordering are possible for a given mother phase. For ice VI more than 40 possible types of ordering were identified.(3) In the early 20th century, Bridgman invented a method to measure pressures up to almost 2 GPa which led him to discover, e.g., the ices V and VI.(4) The large tetragonal $P4_2/nmc$ unit cell of ice VI contains 10 crystallographically different water molecules. Oxygen atoms are arranged in two interpenetrat-ing but not interconnecting networks of hexamer cages. This "self-clathrate" is one of the most complicated of all known crystalline ice phases.(5-7) In the 1960s partial hydrogen order in ice VI was inferred.(5) In 1976 Johari and Whalley observed a slow ordering transformation in ice VI at low temperatures.(8, 9) More than thirty years later Salzmann et al. reported the structure of what we here call ice α-XV, the hydrogen ordered form of ice VI.(10) While theoretical calculations predicted ferroelectric ordering and $Cc$ space group symmetry, experimental data indicate antiferroelectric ordering and $P\bar{1}$.(6, 10, 11) The hydrogen ordering in ice α-XV does not reach completion, but remains partial. In order to explain why only a partial transformation from ice VI to ice α-XV is achieved small ferroelectric $Cc$ nanocrystallites were proposed, which are destabilized as soon as the dielectric constant of the transforming ice VI is reduced.(12, 13) Although ice α-XV was scrutinized very well by different methods(3, 14-16), especially its phase transition behavior is still not fully understood.

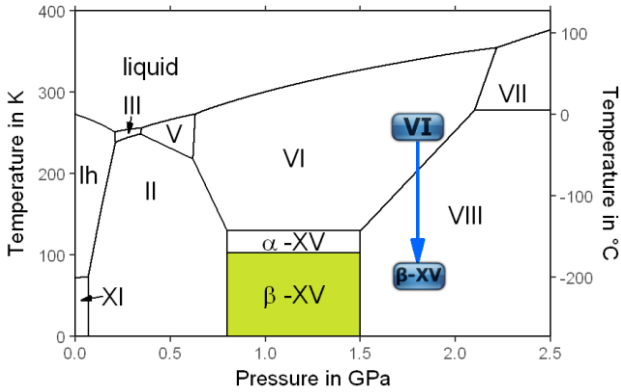

**Figure 1.** The equilibrium phase diagram of H$_2$O including the new hydrogen ordered phase ice β-XV. The blue arrow at 1.8 GPa shows the cooling process of ice VI from 255 K leading to ice β-XV at 77 K, which is metastable in the ice VIII domain.

The complicated nature of ice VI involving two unconnected networks implies that defects migrating in one of them are unable to switch to the other network. This is a source for complex thermal signatures associated with the disordering in ice α-XV. For other ordered ices, such as ice XI(17), ice XIII(18) and ice XIV(19), the thermal signatures of hydrogen disordering just involve a single endotherm. For ice α-XV a weak exotherm preceding the endotherm indicating the first-order transition to ice VI was observed by Shephard and Salzmann.(15) This finding was explained by the interaction between the two disconnected networks, where initial ordering in one of them triggers the disordering in the other.(15) Similarly, intra-network hydrogen ordering takes place upon cooling at temperatures that are higher than those at which inter-network ordering occurs.(15) Therefore, a variation of the original preparation protocol of ice α-XV can affect the ordering scheme of the hydrogen atoms. Based on this insight we studied the influence of preparation pressure and cooling rates on the hydrogen ordering. This led us to the experimental realization of a second hydrogen ordered variant of ice VI, which we call ice β-XV: It is more ordered than the known variant, here called ice α-XV.(10) By heating ice β-XV we observe an unprecedented entropy-driven phase transition between two ordered ice phases. This adds a rectangular stability region to the phase diagram in the GPa pressure range as shown in fig. 1. The newly discovered first transition in the H-atom

ordering at ≈103 K precedes the known second disordering at ≈129 K.(10)

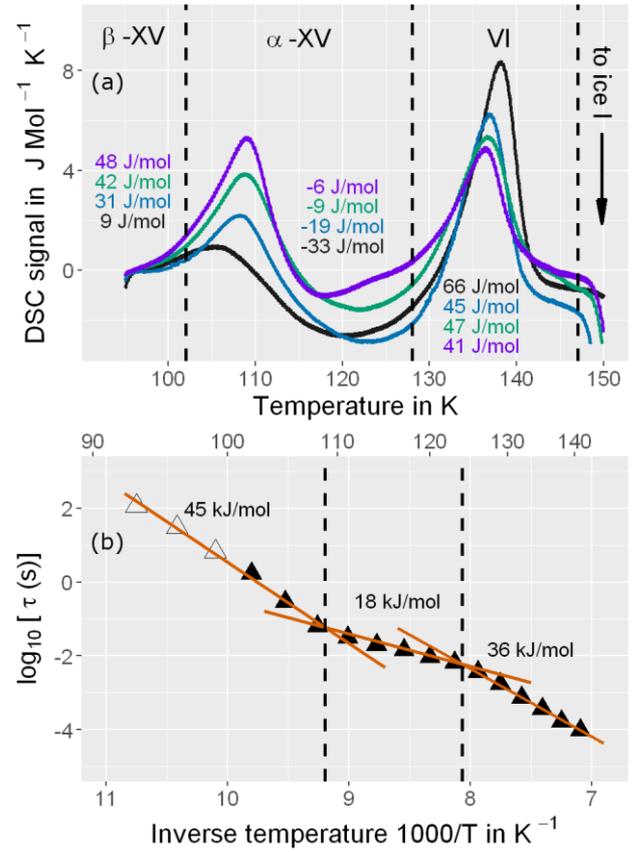

**Figure 2.** (a) Differential scanning calorimetry (DSC) thermograms of ice β-XV samples prepared at pressures of 1.00 GPa (black), 1.45 GPa (blue), 1.60 GPa (green) and 1.80 GPa (violet) and cooled with approximately 3 K min$^{-1}$. The indicated enthalpies are average values from 19 (violet), 4 (green), 5 (blue) and 8 (black) samples. Heating rate is 10 K min$^{-1}$ for all DSC traces. (b) Relaxation map obtained from dielectric loss maxima. Lines are linear fits to the data points. At the ice β-XV to ice α-XV transition the slope of the curve, i.e. the activation energy clearly changes (black numbers).

## Calorimetry and dielectric relaxation experiments

Previously, ice α-XV was prepared by quenching HCl-doped ice VI kept at 1 GPa at rates of about 40 K min$^{-1}$.(15) By reducing the cooling rate at 1.00 GPa to 3 K min$^{-1}$ a weak endotherm of 9 J mol$^{-1}$ appears in calorimetry experiments at 1 bar at an onset temperature of 103 K (bottom trace in fig. 2(a)), that was not noticed for samples cooled at 40 K min$^{-1}$.(15) By increasing the pressure to 1.45, 1.60 and 1.80 GPa this previously unknown endotherm grows to 31, 42 and 48 J mol$^{-1}$, respectively. On the other hand the endotherm with an onset at 129 K indicating the ice α-XV→iceVI disordering transition shrinks in the same pressure range from 66 to 41 J mol$^{-1}$ (see fig. 2(a)). For 1.80 GPa the size of the first endotherm discovered in this work exceeds the size of the previously known second one (top trace in fig. 2(a)). Furthermore, the increase in preparation pressure shifts the ice α-XV→ice VI endotherm up by 3.5 K, but the new endotherm down by 3.9 K. The exotherm noted in ref. 17 between the two endotherms is seen in fig. 2(a) as well, but is comparably weak and plays only a minor role. The data in fig. 2(a) represent an unprecedented observation of two well-separated endotherms, both associated with hydrogen disordering. The configurational entropy changes by 0.46 J K⁻¹mol⁻¹ and 0.33 J K⁻¹mol⁻¹ at the first and the second peak, respectively. This corresponds to 14% and 10% of the Pauling entropy which refers to the maximum entropy change achievable through disordering hydrogen atoms in ice samples. We are not aware of a related two-step disordering neither for $H_2O$ ice nor for related systems such as the spin ices(20-23).

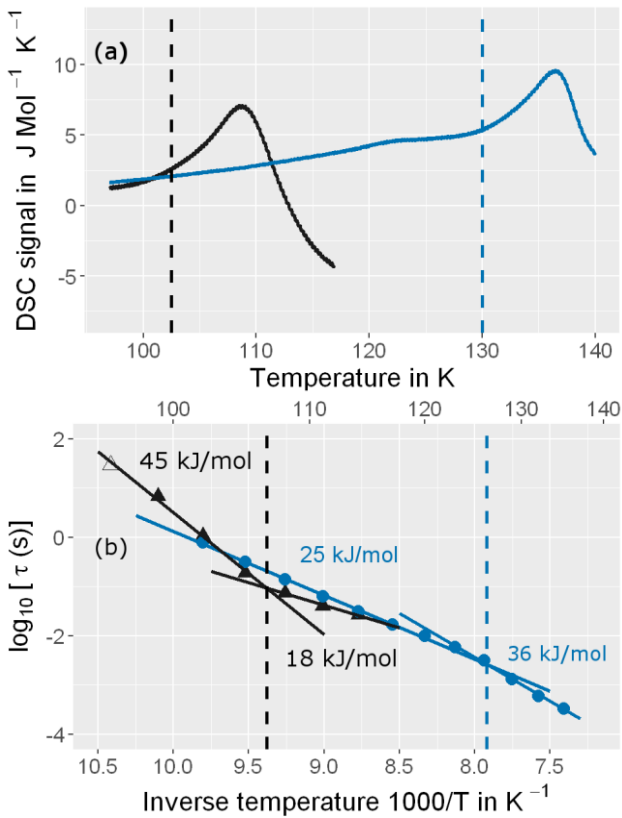

**Figure 3.** Influence of recooling at ambient pressure on ice β-XV. (a) Differential scanning calorimetry (DSC) thermograms of ice β-XV samples prepared at 1.80 GPa heated to 117 K (black trace), recooled to 93 K at 1 bar min⁻¹ and heated a second time to 140 K (blue trace). Heating rate is 10 K min⁻¹ for all DSC traces. (b) Relaxation map obtained from dielectric loss maxima. Ice β-XV was heated to 114 K (black symbols), recooled to 102 K and heated to 135 K (blue symbols). The lines are linear fits to the data points.

At least two interpretations can be invoked to understand the observed phenomena, both of which involve a new hydrogen ordered phase of ice. The first interpretation involves a heterogeneous sample containing both ice β-XV and ice α-XV domains. Upon disordering both domains convert to ice VI, the former at 103 K, the latter at 129 K. The second one involves a two-step disordering in homogeneously ordered crystals according to the sequence ice β-XV→ice α-XV→ice VI. Dielectric loss spectroscopy on the powdered sample clearly indicates three linear regimes in the relaxation map (fig. 2(b)). The transitions take place at 108 K and 124 K, respectively, which favors the latter interpretation. The slopes in the Arrhenius-plot indicate activation energies of 45 kJ mol⁻¹ and 18 kJ mol⁻¹ for ice β-XV and ice α-XV, respectively. These activation energies suggest a stronger hindrance against water dipole reorientation in the more ordered ice β-XV phase as compared to the previously known ice α-XV.

Another difference between the two distinct polymorphs of ice XV is the behavior upon recooling at ambient pressure shown in fig. 3. On first heating to 117 K (black trace in fig. 3(a)) ice β-XV shows the endotherm already seen in fig. 2(a). After recooling from 117 K to 93 K at 1 bar in the calorimeter the first endotherm does not reappear on heating (blue trace in fig. 3(a)). Instead, a flat baseline is seen upon second heating – which implies that ice β-XV does not form from ice α-XV upon cooling at 1 bar. The ice α-XV→ice VI endotherm, however, appears at the same temperature as in fig. 2(a), also after recooling from ice VI at 130 K- which implies that ice α-XV forms from ice VI upon cooling at 1 bar. In the dielectric relaxation map the kink associated with the ice β-XV→α-XV transformation appears on first heating (black triangles in fig. 3(b)). After recooling from 114 K to 102 K this kink no longer appears on second heating (blue circles in fig. 3(b)). Instead only the kink known from fig. 2(b) associated with the ice α - XV→ice VI transformation is seen at around 126K. That is, only two linear regimes and hence two phases are apparent in the relaxation map of the recooled sample, namely ice α-XV and ice VI. Note that the slope of the ice α-XV phase differs slightly between first and second heating (see fig. 3(b)). We attribute the difference to different degrees of order in ice α-XV. The order in ice α-XV is affected in the aftermath of the transition due to the interaction between the networks. This gives rise to the exotherm centered at around 120 K visible in the first heating scan (see fig. 2(a) and 3(a) black trace), which we suggest to call "enthalpy relaxation". This indicates that the transition from ice VI to ice α-XV can be achieved not only at 1 GPa and above, but also at 1 bar. By contrast, ice β-XV does not form from ice α-XV or ice VI at 1 bar, but only at high pressures exceeding 1 GPa. This reinvigorates the notion that the pressure-induced proximity enhances the interaction between the two interpenetrating networks, which might be the key to ice β-XV formation. At high pressures, e.g., 1.8 GPa, the formation of ice β-XV upon cooling and its transformation to ice α-XV are reproducible and reversible. This effect of pressure is also known for the ice XII/XIV phase transition, for which orthorhombic stress needs to be overcome by means of high pressure to achieve the phase transformation.(24)

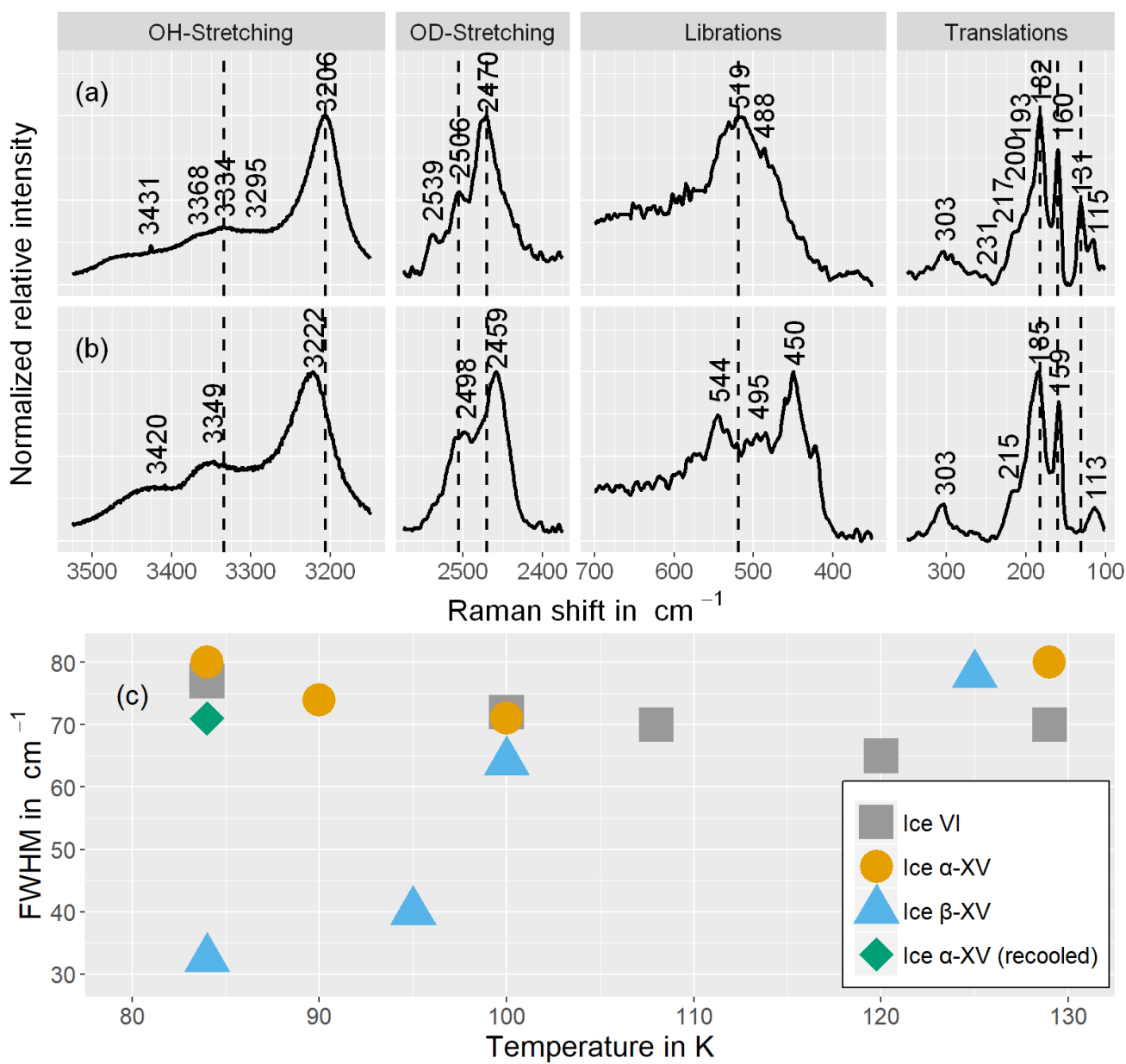

**Figure 4.** Raman spectra of (a) ice β-XV and (b) ice α-XV samples, both containing 9 mol% HDO and recorded at 84K. Frame (c) shows the full width at half maximum of the OD-stretching peak. Ice α-XV (recooled) was heated to 130 K at ambient pressure and recooled to 84 K before the measurement.

## Raman spectroscopy

The nature of the hydrogen ordering was additionally scrutinized using Raman spectroscopy (fig. 4) because this method is sensitive to the local molecular environment. In particular, hydrogen ordering leads to a narrowing and/or splitting of bands.[25] This was shown for instance for the ice XIII/V pair, which features 3-4 separated peaks and shoulders at 100 K in the ordered ice XIII phase, but only a single broad peak for disordered ice V at 110 K.[24] Fig. 4 shows that there are significant shifts in peak positions between ice α-XV and ice β-XV, e.g. the band at 3206 cm$^{-1}$ blueshifts by 16 cm$^{-1}$ while the band at 2470 cm$^{-1}$ redshifts by 11 cm$^{-1}$. Furthermore the peak intensities change massively for selected bands, e.g., the band at 131 cm$^{-1}$ is absent for ice α-XV (panel b) but of medium intensity for ice β-XV (panel a). The three librational bands near 450 cm$^{-1}$, 500 cm$^{-1}$ and 540 cm$^{-1}$ are strong, weak and strong for ice α-XV, but absent, strong and absent for ice β-XV. These findings suggest different selection rules, i.e., different vibrations being symmetry allowed, and hence different space groups for ice β-XV and ice α-XV. It also hints at a pure phase of ice β-XV rather than a mixture of ice phases. Moreover, the full width at half maximum (FWHM) is clearly reduced for all bands in ice β-XV (fig.4, panel a) compared to ice α-XV (fig. 4, panel b) for samples containing 9 mol% of HDO. The broad translational band shows much more features for ice β-XV than for ice α-XV: at least 3 new bands or shoulders are resolved. The FWHM is reduced from 71 to 33 cm$^{-1}$ for the decoupled OD-stretching band. These features are clearly in favor of ice β-XV as a phase distinct from ice α-XV and of a larger degree of hydrogen ordering.

Upon heating ice β-XV one notices a sudden change in the spectra near 100 K to the known ice α-XV spectra, e.g., a jump-like change of the FWHM at 2470 cm$^{-1}$ to the known value for ice α-XV[16] (fig. 4(c)). The transition temperature is in quantitative agreement with the results from other methods, see fig. 2(a). These findings allow one to exclude a scenario in which ice β-XV domains transform directly to ice VI upon heating and demonstrate that the transition sequence involves three phases, a more ordered phase (ice β-XV), a less ordered phase (ice α-XV) and the disordered phase (ice VI). That is, there is a first-order transition between two hydrogen-ordered phases that absorbs latent heat. The endothermic feature near 100K is not a glass transition since, unlike the characteristics expected for the latter phenomenon, it disappears upon reheating.

## Thermodynamic considerations

Based on the calorimetry and Raman results ice β-XV is more ordered and hence thermodynamically favored over ice α-XV at temperatures below 103 K. This means, ice β-XV represents a new stable phase. Its stability range in the phase diagram is shown in green in fig. 1. According to the Clapeyron relation $\frac{dp}{dT} = \frac{\Delta S_m}{\Delta V_m}$ the phase boundaries between the ice β-XV/ice II and the ice β-XV/ice VIII neighbors have to be almost vertical. This is because of the small change in entropy occurring at the phase transition between two ordered phases of differing densities. Furthermore, the densities of ice α-XV and ice β-XV are similar, as inferred from the X-ray experiments that we present below. Therefore, the phase boundary between these two phases needs to be almost horizontal which is supported by the experimental data. Hence, an almost rectangular stability domain results. Thus, two triple points, each associated to three hydrogen ordered phases, are identified at $T_{II-αXV-βXV}$ = 103K, $p_{II-αXV-βXV}$ = 0.8 GPa and at $T_{VIII-αXV-βXV}$ = 103K, $p_{VIII-αXV-βXV}$ = 1.5 GPa.

## Powder X-ray diffraction

In order to learn more about the nature of the hydrogen order in ice β-XV, in particular whether or not it displays ferroelectricity, neutron diffraction experiments on deuterated samples would be desirable. However, a second endotherm, analogous to that observed in fig. 2(a), does not appear in fully deuterated samples, i.e., a pronounced isotope effect precludes such experiments.[3, 14] Specifically a DCl-doped $D_2O$ sample cooled at 1.4 GPa at 0.5 K min$^{-1}$ does not show the second endotherm. This is consistent with the neutron diffraction results obtained by Salzmann et al.[10]: A deuterated sample prepared the way just described did not show any indications of ice β-XV but rather the signatures of ice α-XV. Also Komatsu et al. did not report any hints for the existence of a phase other than ice VI and ice α-XV in deuterated samples, even though the pressure range up to 1.6 GPa was investigated.[3] For a crystallographic comparison of ice β-XV and ice α-XV we used powder X-ray diffraction on $H_2O$ samples rather than neutron diffraction on $D_2O$ ices. Hydrogen atoms barely scatter X-rays. Therefore, different types of hydrogen ordering can merely be inferred from indirect effects such as slight changes in the oxygen network, as for instance exploited by Kamb et al. for the example of ice II.(26)

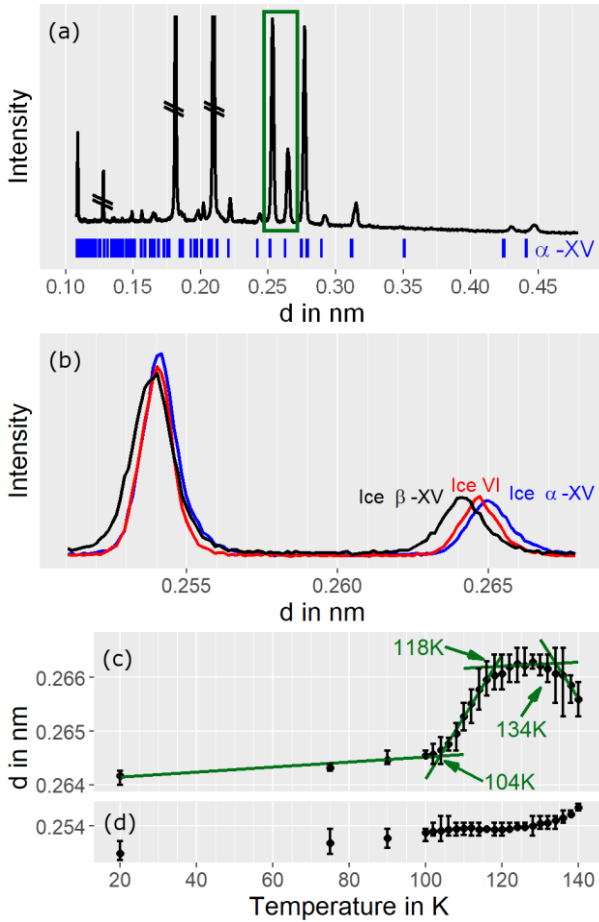

**Figure 5.** Panel (a) shows the powder X-ray diffractograms of ice β-XV recorded at 20K. Panel (b) a comparison of diffraction pattern for ice VI, ice α-XV, and ice β-XV at 80K in the range highlighted by the green rectangle in (a). Peak positions obtained at different temperatures are shown in panel (c) and (d).

Indeed, the X-ray diffractogram in fig. 5(a) for ice β-XV recorded at 20 K is very similar to the known X-ray diffractograms of ice VI(27) and ice α-XV(10) (see ticks). This demonstrates that the oxygen network in ice β-XV is almost identical to the ones in ice VI and ice α-XV. Hence, both endotherms in fig. 2(a) are indeed associated with sequential hydrogen disordering rather than to significant changes in the oxygen network. The most important difference, namely the Bragg peak near d=0.265 nm, is magnified in fig. 5(b). In antiferroelectric ice α-XV this peak is shifted to higher d-spacing compared to ice VI whereas in ice β-XV it is shifted to lower d-spacing .(14) We surmise this to be caused by a ferroelectric nature of the new phase, but need yet to determine its space group and unit cell. The temperature dependence of the lattice spacings (fig. 5(c)) is quite revealing: At 104K (arrow in fig. 5(c)) there is a pronounced increase in the lattice plane spacing marking the first peak in fig. 2(a). At 118K the expansivity rapidly decreases, marking the onset of the weak exotherm and transition to ice α-XV in fig. 2(a). Finally, the contraction observed at around 134 K marks the second endotherm in fig. 2(a), i.e., the transition to ice VI. Other Bragg peaks, e.g., the one near d=0.254 nm (fig. 5(d)) do not show this phenomenology, as expected for an interaction caused by a directional electric field of ordered hydrogen atoms.(14, 15) This rules out that ice β-XV and ice α-XV are identical and again demonstrates the sequential presence of three distinct phases upon heating, namely ice β-XV, ice α-XV and ice VI.

**Conclusions**

In summary, we found evidence for the existence of a second ordered phase of ice VI. Based on Raman shifts, the appearance of new bands in ice β-XV and intensity changes between two phases alone, strong indications for differences in their space groups are found. The observation of a discontinuous transition involving latent heat in fig. 2(a) as well as the sudden change of activation energies indicated in the relaxation map of fig. 2(b) underscores that ice β-XV and ice α-XV are to be understood as distinct phases. It remains to be determined whether ice β-XV represents the ferroelectric *Cc* phase originally envisaged for ice α-XV which is now considered to be antiferroelectric. In principle, the crystal structure of ice β-XV might show any of the 45 possible types of dipole ordering detailed by Komatsu et al.(3) Knowledge of the crystal structure based on a diffraction experiment is traditionally required to give a unique roman numeral to a new ice phase.(28, 29) Since we do not have this information at hand we call the second ordered variant of ice VI as ice β-XV. The previously known ordered variant of ice VI, called ice XV in literature(10), is here referred to as ice α-XV. We recommend to discard this α/β designation as soon as the crystal structure is published and to replace ice β-XV by "ice XVIII". Furthermore, we established its stability domain in the phase diagram of water and clarified that it experiences two first-order phase transitions: first the unprecedented order-order transition to ice α-XV and then the order-disorder transition to ice VI. In terms of Pauling entropy[30] ice β-XV is approximately 14 % more ordered than ice α-XV, as judged from the size of the endotherm in fig.2(a). The experiments reported in this work make the case for a second hydrogen ordered phase related to a disordered ice phase and the possibility of governing the electric properties of ice samples by their preparation history.

## Methods

**Preparation of ice samples** Ice α-XV was prepared following the protocol by Shephard and Salzmann (2015)[15], namely by cooling 600μl 0.01 M HCl in $H_2O$ (in case of ice VI pure $H_2O$) to 77 K, compressing to 1.0 GPa using ZWICK BZ100 /TL35 material testing machine with a piston-cylinder setup (8 mm diameter bore),heating to 255 K and quenching with liquid nitrogen at ≈45 K min⁻¹ to obtain ice α-XV. Ice β-XV was prepared by using a variation of this protocol. After cooling the sample to 77K, pressures of 1.0 - 1.8 GPa were applied before heating it to 255K. The sample was then isobarically cooled at ≈3K min⁻¹ to recover ice β-XV. By heating ice β-XV at 1.8 GPa to 255K and recooling it to 77K at a rate of ≈3K min⁻¹, the reproducibility of the phase transition under pressure was probed.

**Differential scanning calorimetry** Ice β-XV as well as ice α-XV were scrutinized calorimetrically using a Perkin Elmer DSC 8000 and compared with earlier works of Shephard and Salzmann.[15] The samples were filled into aluminum crucibles under liquid nitrogen and directly transferred to the precooled oven of the calorimeter. Every sample was characterized by heating at 10 K min⁻¹ from 93 K to 253 K – thereby observing the transition sequence [ice β-XV→]ice α-XV→ice VI→ice I.[15, 31] To determine the sample mass used for the measurement, the endothermic melting peak at 273 K was compared with the known heat of fusion of water, 6012 J mol⁻¹.[32] Additionally annealing experiments[15, 16] at 117 K were conducted to probe the reversibility of the order→order phase transition. To this end the respective sample was heated to the annealing temperature inside the DSC oven, was kept at this temperature for 30 – 60 minutes, recooled to 93K and heated at 10 K min⁻¹.

**Dielectric relaxation experiments** To measure the dielectric response of the ice crystals, the samples were transferred from Innsbruck to Dortmund under liquid nitrogen conditions. An Novocontrol Alpha-A impedance analyzer was used to investigate the dielectric response as described in earlier works.[19, 33] The powdered crystals were cold-loaded into a parallel plate capacitor connected to a Quatro cryosystem. An exact determination of the cell´s filling factor was not possible.. Starting at 96 K, the samples were heated at average rates of 0.1 K min⁻¹ – 0.6 K min⁻¹ to 140K (fig. 2(b)) or first heated to 114 K, recooled to 102 K and heated a second time to 135 K. Closed symbols in the Arrhenius plots represent relaxation times, which were determined by fitting the dielectric loss peak using a Havriliak-Negami function. Open symbols were determined by frequency-temperature-superposition.

**Raman Spectroscopy** All Raman spectra were recorded with a WiTec WMT50 Spectrometer (532nm laser, 20 mW) inside an Oxford N Microstat controlled by a Lakeshore 331S temperature controller. For obtaining the spectra in fig.4(a) ice β-XV, prepared at 1.8 GPa, was powdered and loaded into the microstat at 84 K. Crystals were selected under the microscope for Raman measurements in confocal geometry. For obtaining the spectra in fig.4(b) ice α-XV, prepared at 1.0 GPa, was heated to 130 K in the Oxford N Microstat and recooled to 84 K as described by Whale et al.[16]. Raman spectra were taken, using a 1800 mm⁻¹ grating and accumulation times of about 30 minutes. The data are smoothed with a 2nd-order Savitzky-Golay filter (7 points). For fig 4(c) temperature dependent Raman spectra were recorded for ice VI, ice β-XV and ice α-XV. One additional point is shown for an ice α-XV sample recooled from 130 K as described in ref. 16.

**Powder X-ray diffraction** For powder X-ray diffraction measurements samples were powdered under liquid nitrogen and transferred to the sample chamber at 20 K to 80 K. The pattern in fig. 5(b) was recorded in flat geometry on a Cu-Ni sample holder with a Siemens D5000 diffractometer. The peak intensities were scaled to match at 0.265 nm. The other diffractograms were recorded with a XPERT III PANalytical Xpert Pro MPD diffractometer (crossed out peaks in panel (a) stem from the sample holder). All diffractograms were obtained at a Cu Kα x-ray source in θ-θ Bragg-Brentano geometry and by using a Cu sample holder.

**Acknowledgements** We thank Werner Artner, Fabian Weiss, and Klaudia Hradil of the X-ray Center of the Technische Universität Wien for providing the XRD instrument and for their help during the measurements. We are also grateful to the Austrian Science Fund (FWF) for the financial support (Project I 1392) and to the Deutsche Forschungsgemeinschaft (Grant No BO1301).